# Origin of Polar Distortion in LiNbO$_3$–type "Ferroelectric" Metals: Role of A-site Instability and Short-Range Interactions

H. J. Xiang

Key Laboratory of Computational Physical Sciences (Ministry of Education), State Key Laboratory of Surface Physics, and Department of Physics, Fudan University, Shanghai 200433, P. R. China

e-mail: hxiang@fudan.edu.cn

**Abstract**

Since conduction electrons of a metal screen effectively the local electric dipole moments, it was widely believed that the ferroelectric-like distortion cannot occur in metals. Recently, metallic LiOsO$_3$, was discovered to be the first clear-cut example of an Anderson-Blount "ferroelectric" metal, which at 140 K undergoes a ferroelectric-like structural transition similar to insulating LiNbO$_3$. This is very surprising because the mechanisms for structural phase transitions are usually quite distinct in metals and insulators. Through performing first principles calculations, here we reveal that the local polar distortion in LiOsO$_3$ is solely due to the instability of the A-site Li atom, in contrast to the LiNbO$_3$ case where the second order Jahn-Teller effect of the B-site Nb ion also plays an additional role. More importantly, the "ferroelectric"-like long range order of the local polar distortion is found to be due to the predominantly ferroelectric short-range pair interactions between the local polar modes which are not screened by conduction electrons. Furthermore, we predict that LiNbO$_3$–type MgReO$_3$ is also a "ferroelectric" metal, but with a much higher structural transition temperature by 391 K than LiOsO$_3$. Our work not only unravels the origin of FE-like distortion in LiNbO$_3$–type "ferroelectric" metals, but also provides clue for designing other multi-functional "ferroelectric" metals.

**Introduction**

Although metallic behavior and ferroelectricity have long been thought to be incompatible because conduction electrons screen the internal electric field, Anderson and Blount[1] nonetheless suggested almost half a century ago that certain structural transitions observed in metals might be "ferroelectric" (FE) in nature. To be considered as an Anderson-Blount "FE" metal, several criteria have to be met: The structural transition has to be continuous, and the low temperature structure has to be non-centrosymmetric and FE-like. Recently, $LiOsO_3$ was identified as the first clear-cut example[2] of the "FE" metal. Similar to insulating $LiNbO_3$[3], metallic $LiOsO_3$ displays the centrosymmetric $R\bar{3}c$ rhombohedral structure (See Fig. 1a) at high temperatures and transforms to the FE-like $R3c$ structure (See Fig. 1b) at 140 K, involving a shift in the position of the Li atoms. Given the metallic nature of $LiOsO_3$, the observation of FE-like phase transition in this system is quite surprising: free electrons typically screen the long-range dipole-dipole interactions that favor the off-center displacements, possibly making ferroelectricity unstable. The microscopic mechanism for the FE-like distortion in $LiOsO_3$ remains puzzling. Even for $LiNbO_3$, there are two opposite interpretations of the role of the Li ion in the mechanism for the polar distortion. In one interpretation it is the driving mechanism (active role), and in the other it is not (passive role). The former examples include the Li-ion related displacement FE theory proposed by Lines[4] and the order-disorder process involving the Li ions considered by Safaryan[5]. On the other hand, Inbar and Cohen[6] proposed the latter scenario: The driving mechanism behind the FE instability in $LiNbO_3$ is the hybridization between the Nb 4d states and the O 2p states, which leads to the displacements of the O ions towards the Nb ions; these O displacements are responsible for the lithium displacements from their centrosymmetric sites, indicating that the Li ions themselves are passive players in the FE transition. Because of the structural similarity between LiOsO$_3$ and LiNbO$_3$, the clarification of the mechanism for the FE-like transition in LiOsO$_3$ will also shed new light on the transition mechanism in the prototypical FE compound LiNbO$_3$.

In this work, we investigate the microscopic mechanism for the FE-like transition in LiOsO$_3$ by performing density functional theory (DFT) calculations and Monte Carlo (MC) simulations using a first-principles-based effective Hamiltonian[7]. We find that the under-coordinated A-site Li atom is responsible for the FE-like instability in LiOsO$_3$: The Li-atom tends to displace toward the neighboring out-of-plane O atoms to lower the electrostatic energy. This is different from the case of LiNbO$_3$, where both A-site Li and B-site Nb ions appear to be relevant to the FE transition. By constructing an effective Hamiltonian for the local polar modes, we find that there exist significant interactions between the local polar modes despite of the presence of Os 5d conducting electrons at the Fermi level. In particular, two short-range pair interactions are strongly FE-like in order to lower the electrostatic energy and/or make the O displacement cooperative. Finally, the understanding of the FE-like transition in LiOsO$_3$ helps us to predict a new room temperature "FE" metal, .i.e, LiNbO$_3$–type MgReO$_3$.

**Computational details**

Total energy calculations are based on DFT within the generalized gradient approximation (i.e., PBE)[8] on the basis of the projector augmented wave method[9] encoded in the Vienna ab initio simulation package[10]. We also check that the main results will not change if a different density functional [such as local density approximation (LDA) or PBEsol[11]] is adopted. For example, the energy differences between $R\bar{3}c$ and $R3c$ LiOsO$_3$ are 0.042 eV per formula unit (f.u.) and 0.028 eV/f.u. when adopting LDA and PBEsol, respectively. These results are close to the PBE value (0.021 eV/f.u.). Unless otherwise noted, the plane-wave cutoff energy is set to 500 eV. A 12×12×12 k-mesh for the rhombohedral unit cell is adopted for the Brillouin zone sampling.

To understand the interaction between the local polar modes, we employ an effective Hamiltonian, which includes the local mode self energy term $E^{self}$ and the short-range pair interactions between the local modes $E^{short}$: $E^{tot} = E^{self}(\{u\}) + E^{short}(\{u\})$. Following the original work by Zhong *et al.*[7], the self energy is defined as: $E^{self}(\{u\}) = \sum_i \sum_{m=1}^{2} E(u_{i,m})$, where $E(u_{i,m})$ is the energy of the m-th isolated local mode at cell $R_i$ with amplitude $u_{i,m}$, relative to that of the $R\bar{3}c$ structure. $E(u_{i,m})$ contains harmonic and the fourth order anharmonic contributions: $E(u_{i,m}) = k_2 u_{i,m}^2 + \alpha u_{i,m}^4$. For the short-range interactions $E^{short}$, they stem from differences of the short-range repulsion and electronic hybridization between two adjacent local modes and two isolated local modes. As a second order approximation, $E^{short}$ can be written as: $E^{short}(\{u\}) = \frac{1}{2} \sum_{(i,m) \neq (j,n)} J_{im,jn} u_{im} u_{jn}$, where $J_{im,jn}$ is the coupling constant between the local mode $u_{im}$ and $u_{jn}$.

To evaluate the parameters of the effective Hamiltonian of the local mode, we use a large supercell (i.e., 320-atom cell) so that the interaction between the local mode and its periodic images of the local mode is negligible. Here, a 4×4×4 k-mesh is adopted for the Brillouin zone sampling. To obtain of $k_2$ and $\alpha$ of the self energy term, the magnitude of one local mode is set to several different values (from 0 to 0.8 Å), while the magnitude of all the other local modes in the supercell are 0. For the coupling constants J of the short-range pair interaction between the local modes, we consider two states: (i) FE-like state: Both magnitude of the local modes are set to a reasonable positive value d (e.g., 0.4 Å); (ii) AFE-like state: The magnitude of the two local modes are set to d and $-d$ Å, respectively. In both cases, the magnitude of all the other local modes in the supercell are fixed to 0. The coupling constant J can then be estimated from the energy difference between the two states as: $J = \dfrac{E(FE) - E(AFE)}{2d^2}$.

The parallel tempering Monte Carlo (PTMC) method[12] is adopted to study the thermodynamic behavior of the structural phase transition in $LiOsO_3$. Parallel tempering, also known as replica exchange, is a simulation method aimed at improving the dynamic properties of the usual Markov chain MC sampling method. Essentially, one runs several copies of the system at different temperatures. Then, based on the Metropolis criterion one exchanges configurations at different temperatures. The idea of this method is to make configurations at high temperatures available to the simulations at low temperatures and vice versa. In our PTMC simulations of the effective Hamiltonian, a $10 \times 10 \times 6$ supercell of the hexagonal unit cell is adopted. Our test shows that the results obtained from a $12 \times 12 \times 8$ supercell are very similar. The number of replicas is set to 96.

**Results and discussions**

We find that it is useful to define a tolerance factor ($t_R$) for describing the polar instability in a $R\bar{3}c$ structure. In the centrosymmetric $R\bar{3}c$ structure, for each A-site ion, there are three nearest neighbor in-plane O ions and the other six further neighboring out-of-plane O ions. As shown in Fig. 1c, the A-site ion is almost closely packed in center of in-plane O triangle so that the distance ($r_{AO}$) between the A-site ion and the O ion equals approximately the sum of the ionic radii[13] (6-coordinated) of the A-site ion and O ion ($r_{AO} \approx r_A + r_O$). The B-site ion sits in the center of the oxygen octahedron with the distance ($r_{BO}$) between the B-site ion and O ion close to the sum of the ionic radii of the B-site ion and O ion ($r_{BO} \approx r_B + r_O$). If the oxygen octahedron is perfect, i.e., the <O-B-O angle $\alpha$ is $90°$, we can prove from geometrical consideration that the distance ($R$) between the A-site ion and out-of-plane O ion satisfies $R = \sqrt{2} r_{BO}$. The tolerance factor can be defined as $t_R = \dfrac{r_A + r_O}{R} \approx \dfrac{r_A + r_O}{\sqrt{2}(r_B + r_O)}$. Note that we do not assume an ideal rhombohedral angle (60°) in the derivation of the tolerance factor. Although the tolerance factor for the $R\bar{3}c$-type structure defined here turns out to be the same as the Goldschmidt tolerance factor ($t_P$) for describing the perovskite structure[14], the starting geometrical structures are different: $t_R$ and $t_P$ are derived by considering the $R\bar{3}c$ and $Pm3m$ (cubic perovskite) structures, respectively. These two tolerance factors also apply to different situations: $t_P$ is used to describe the instability of a perfect cubic perovskite structure, while $t_R$ is suitable for describing the polar instability of a $R\bar{3}c$ structure, which is already severely distorted from the cubic perovskite structure. In our definition of the tolerance factor for the $R\bar{3}c$ structure, we consider the distance between the A atom and the out-of-plane O atom, which is different from the distance between the A atom and the in-plane O atom due to the octahedron rotation. In contrast, Goldsmith considered the distance between the A atom and an O atom (note that all the distances between the A atom and the twelve neighboring O atoms are equal) when he defined the tolerance factor for the cubic perovskite structure. It should be noted that the strong correlation between tolerance factor $t_P$ for describing the perovskite structure and ferroelectric mode frequency in the $R\bar{3}c$ structure was already pointed out[15]. When $t_R < 1$, the A-site ion is too far from the out-of-plane O ion, the A-site ion tends to move towards three of the six out-of-plane O ions in order to decrease the electrostatic energy or form chemical bonds, thus there will be an A-site driven polar instability. When $t_R > 1$, the A-site ion will be too close to the out-of-plane O ions if the B-site ion is still in close contact with the six neighboring O ions. In this case, the oxygen octahedron might expand so that the B-site ion might become unstable. Note that the radius of an atom in a metallic system decreases with respect to similar insulating materials. However, the tolerance factor based on Shannon's ionic radii still gives an estimation of the A-site instability in the $R\bar{3}c$ compound.

To see a general trend of polar instability, we compute the energy difference between the $R\bar{3}c$ and $R3c$ structures for a series of materials related to LiOsO$_3$ and LiNbO$_3$. For comparison, we also consider some compounds related to the synthesized LiNbO$_3$-type ZnSnO$_3$[16]. In order to examine the dependence of the polar instability on the ionic radius, we plot [see Fig. 2 (a)] the energy difference versus a tolerance factor ($t_R$) as defined above. Several interesting trends emerge: (i) For ZnSnO$_3$ related compounds (ZnBO$_3$ or ASnO$_3$) with $t_R < 1$, the energy difference between the $R\bar{3}c$ and $R3c$ structures decrease with $t_R$, suggested that the A-site ion is responsible for the polar instability in agreement with our above discussion on the tolerance factor. (ii) For LiOsO$_3$ related AOsO$_3$ (A: alkali metal) systems, the increase of the size of A-site ion makes the $R3c$ structure less stable, as in the case of ASnO$_3$. Our DFT calculations show that the $R3c$ LiOsO$_3$ is more stable by 0.021 eV/f.u. than $R\bar{3}c$ LiOsO$_3$, in agreement with the FE-like phase transition observed experimentally[2]. (iii) The relative stability between $R3c$ and $R\bar{3}c$ phase of ANbO$_3$ systems does not depends monotonically on the size of alkali element A. If $t_R < 1$, the $R3c$ structure becomes less stable as the size of A-site ion increase, in agreement with the usual trend found for the ZnSnO$_3$ and LiOsO$_3$ series. For RbNbO$_3$ and CsNbO$_3$, the tolerance factor is larger than 1 and the $R3c$ structure becomes stable again. This is because the large size of A-site ion make the Nb-O bond length larger than the common Nb-O bond length, thus the Nb ion becomes unstable due to the tendency to hybridize with the neighboring O ions. This is similar to the FE mechanism[17] in perovskite BaTiO$_3$. (iv) From LiTaO$_3$ to LiNbO$_3$, then to LiVO$_3$, the $R3c$ structure becomes more stable although the tolerance factors in these three structures are similar. This may be explained as follows: The band gaps (3.15, 2.47, 0.61 eV for LiTaO$_3$, LiNbO$_3$, LiVO$_3$, respectively from the PBE calculations) of the $R\bar{3}c$ structure increase with the size of the B-site ion, thus the second-order Jahn-Teller effect[18] (hybridization between empty d state of B ion and occupied 2p state of O ion) in small B ion systems becomes more important. It should be noted that some compounds mentioned above may be hypothetical, but are merely used to see the general trend of polar instability of the $R\bar{3}c$ structure.

We also compute the energy difference between $R\bar{3}c$ and $R3c$ structures for LiBO$_3$ systems (some may be hypothetical, but considered to show the trend) where B is the metal element in the same period (No. 6) as Os. As can be seen from Fig. 2(b), the $R3c$ structure is more stable than the $R\bar{3}c$ one for all considered LiBO$_3$ system. This is consistent with our above claim that the A-site Li atom is responsible for the FE-like instability. However, the relative stability of the $R3c$ structure depends on the nominal number of 5d electrons of the B$^{5+}$ ion. It turns out that if the B$^{5+}$ ion has zero 5d electrons (Ta$^{5+}$), six 5d electrons (Au$^{5+}$), or full occupied 5d electrons (Bi$^{5+}$), the stability of the $R3c$ structure is higher than its neighboring systems. All these three systems are insulating. LiTaO$_3$ and LiBiO$_3$ are insulating because the 5d orbitals are either empty or fully occupied. In insulating LiAuO$_3$, the 5d t$_{2g}$ states are fully occupied by six electrons, while the high lying e$_g$ states are empty. In general, the $R3c$ structure is more stable when the LiBO$_3$ system is insulating. This is because the BO$_6$ octahedron is more easily deformed in this case. In particular, the BO$_6$ octahedron with partially occupied t$_{2g}$ states on B$^{5+}$ ion tends to be a perfect octahedron, thus the energy difference between the $R\bar{3}c$ and $R3c$ structures is small.

Our above analysis suggests that the FE-like instability in LiOsO$_3$ is due to the A-site Li atom, similar to the case of ZnSnO$_3$; while both A-site and B-site ions are relevant to the FE displacement in LiNbO$_3$. To elaborate this point more clearly, we compute the phonon dispersion of the $R\bar{3}c$ structure of LiOsO$_3$ and LiNbO$_3$. Fig. S2 of the Supplementary Material[19] shows that there are two isolated imaginary phonon modes throughout the Brillouin zone in LiOsO$_3$. As shown in Fig. 1a, the lowest phonon mode at $\Gamma$ (A$_{2u}$ symmetry) is FE-like, which is mainly dominated by the displacement of the Li atom with small contribution from the O atoms. The displacement of the Os atom in this mode is vanishingly small. In contrast, there is some contribution from the Nb ion to the lowest phonon mode at $\Gamma$ in LiNbO$_3$. Fig. 3 shows the total energy as a function of magnitude of the A$_{2u}$ mode for LiNbO$_3$ and LiOsO$_3$. Instead of the displacement vector from $R\bar{3}c$ structure to $R3c$ structure, the FE-like phonon mode is used as the full mode. In the full mode, the displacements of all atoms are taken into account. We can clearly see the double wells in both systems. To separate the role played by Li and B-site ion, we also show the energy versus the partial modes, i.e., the B-O mode and Li-O mode. The B-O (Li-O) partial mode is obtained by setting the Li-ion (B-ion) displacement in the full A$_{2u}$ mode vector to zero. In LiOsO$_3$, there is no double well for the Os-O mode, but the curve for the Li-O mode is almost the same as the full phonon mode. This suggests that the Os atom is not unstable in LiOsO$_3$. In contrast, there are double wells for both the Nb-O mode and the Li-O mode, although the double well in the former case is shallower. Thus, the phonon mode analysis supports the idea that both A-site and B-site ions contribute to the FE instability in LiNbO$_3$, but the Li atom instead of the Os atom is responsible for the FE-like instability in LiOsO$_3$. The key difference between LiNbO$_3$ and LiOsO$_3$ is the occupation of the d states: The 4d states of the Nb$^{5+}$ ion are empty, thus may lower the total energy by hybridizing with the occupied O 2p states, facilitating the polar distortion; since the 5d t$_{2g}$ states of the Os$^{5+}$ ion is partially occupied, the second order Jahn-Teller effect could not take place.

We have shown above that the Li atom in $R\bar{3}c$ LiOsO$_3$ is unstable because it is too far from the out-of-plane O atoms. For each Li atom, it has two choices for the displacement, either move up or move down. If there is no interaction between the Li displacements, the Li positions will be disorder, thus overall the system will be non-polar. The key question is how the Li displacements interact with each other to realize a FE-like state in LiOsO$_3$. To address this issue, we construct an effective Hamiltonian (see **Computational details**) to describe the interaction between the Li displacements. The phonon dispersion (see Fig. S2) shows that there exist only two imaginary phonon modes and these two modes are separated from the other positive phonon modes. Therefore, it is reasonable to describe the low-energy behavior of LiOsO$_3$ by considering only these two phonon modes. Each unit cell of the $R\bar{3}c$ structure contains two Li atom. In the lowest phonon mode at Γ, two Li atoms displace along the same out-of-plane direction. In contrast, two Li atoms move opposite to each other in the other higher imaginary phonon mode at Γ. By combining the two imaginary phonon modes at Γ, we obtain the local phonon mode for each of the two Li atoms. As can be seen in Fig. 4, the local mode is dominated by displacement of the Li atom at the center of the local mode along the c axis. The three in-plane O atoms displace a little almost along the negative c-axis, while the upper (lower) three out-of-plane O atoms move towards (opposite to) the Li atom.

We consider the short-range pair interaction up to fifth nearest neighbor (NN) (see Fig. S3). Our test calculation shows that the longer range interactions are much weaker. There is no need to include the dipole-dipole interaction term since LiOsO$_3$ is confirmed to be metallic (see Fig. S1 of Supplementary Material[19]) and the long range dipole-dipole interaction is expected to be fully screened in a metal, while the short-range electrostatic interaction can be described accurately by $E^{short}$. For simplicity, this Hamiltonian does not take into account the strain related degrees of freedom, which should not change qualitatively the nature of interactions between the local modes. The expansion parameters ($k_2$, $\alpha$, and $J$) determined from first-principles calculations are listed in Table I. Negative $k_2$ and positive $\alpha$ will lead to the expected double well potential for the local mode. We find that J$_1$ and J$_3$ are dominant and FE-like. As can be seen from Fig. 4b, J$_1$ is FE-like because the opposite displacements of two nearest neighbor Li atoms will make these Li atoms too close to each other, thus the repulsion between the Li$^+$ ions disfavors the antiferroelectric (AFE) alignment of local modes. The reason why J$_3$ is strongly FE-like is different (see Fig. 4b): These two local modes share an out-of-plane O atom. If the two local modes displace along opposite directions, the shared O atom cannot move to lower the Li-O electrostatic energy, which makes the AFE-like alignment of local modes unfavorable. The coupling constant J$_2$ is weakly AFE-like because the electrostatic interaction energy is favorable when the dipoles of the local mode pair are antiparallel aligned.

Although each individual coupling constant J is smaller than the second order parameter $k_2$ of the self energy term, the pair interactions between the local modes play an important role in stabilizing the FE-like $R3c$ structure. If we consider only the self energy, the lowest total energy is $E^{tot} = -\frac{k_2^2}{4\alpha} = -0.0022$ eV/f.u. when $|u_{i,m}| = \sqrt{-\frac{k_2}{2\alpha}} = 0.239$ Å. When the short-range pair interactions are taken into account, the ground state is the FE-like $R3c$ structure with the total energy $E^{tot} = -\frac{k_{eff}^2}{4\alpha} = -0.0173$ eV/f.u. and $u_{i,m} = \sqrt{-\frac{k_{eff}}{2\alpha}} = 0.401$ Å or $u_{i,m} = -\sqrt{-\frac{k_{eff}}{2\alpha}} = -0.401$ Å, where $k_{eff} = k_2 + 3J_1 + 3J_2 + 3J_3 + 3J_4 + J_5 = -0.2144$ eV/Å. Thus, the short range interactions between the local dipole modes not only makes the long range FE-like arrangement of the dipoles possible, but also greatly stabilize the Li displacement. Our study shows that the short range interactions between the local dipole modes is not screened in a metallic system, and the short range interactions play an important role in stabilizing the FE-like $R3c$ structure. This is different from the case of a conventional FE material (i.e., BaTiO$_3$). As shown by Zhong *et al.*,[7] it is the long-range dipole-dipole interaction which makes the ferroelectric state favorable. If the dipole-dipole interaction is neglected, the ferroelectric instability disappears. The short-range interactions are not responsible for the ferroelectric instability in BaTiO$_3$ since the strongest short-range interaction is even AFE.

Based on the effective Hamiltonian, the finite temperature behavior and the detailed phase transition characteristics in LiOsO$_3$ can be examined by the MC simulations. In the simulation, each unit cell contains two continuous degrees of freedom corresponding to the two local modes. The parallel tempering (PT) technique[12,20] is adopted since it gives better dynamic properties of physical systems than the conventional Metropolis MC sampling method. The low-temperature results from the PTMC simulation confirm that the ground state of LiOsO$_3$ is indeed FE-like with all the local mode aligned along the same direction. The specific heat displays a characteristic single $\lambda$-peak, in agreement with the experimental observation[2]. We note that the phase transition temperature from our simulation is 214 K, which is higher than the experimental value (140 K). This may be due to the error in the density functional or the neglecting of the elastic degrees of freedom. We find that the average local mode can be used as the order parameter for the phase transition: $\langle u \rangle = \sum_{i,m} u_{i,m} / N$, where N is the number of local modes in the simulation cell. The average local mode increase continuously from zero when the temperature decreases, in accord with the nature of the second order transition, as required in a "FE" metal. The saturated zero-temperature value of $\langle u \rangle$ is about 0.4 Å, in excellent agreement with our above analytic derivation.

Although LiOsO$_3$ was identified as the first example of an Anderson-Blount "FE" metal[1], at room temperature it is centrosymmetric because the phase transition temperature (140 K) is lower than room temperature. An interesting question is whether there exists room temperature "FE" metal. Our study reveals that the FE-like instability in LiOsO$_3$ is due to the tendency of Li atom to form bonds with out-of-plane O atoms. To make the $R3c$ structure more stable, one way is to reduce the tolerance factor. Another way is to put a high valence ion on the A-site. If we replace Li$^+$ ion by high valence ion (such as Mg$^{2+}$ ion), the high valence will make the Mg$^{2+}$ ion in the $R\bar{3}c$ structure even more unstable. Our DFT calculation indeed confirms this idea: For MgReO$_3$, the $R3c$ structure is more stable by 0.195 eV/f.u. than the $R\bar{3}c$ structure (We also replace Os by Re because Os$^{5+}$ has the same number of d electrons as Re$^{4+}$), which is much larger than the energy difference in the case of LiOsO$_3$. There are two reasons to account for why $R\bar{3}c$ MgReO$_3$ has a stronger instability than $R\bar{3}c$ LiOsO$_3$: first, $t_R$ for MgReO$_3$ is slightly smaller than that for LiOsO$_3$; second, the energy gain in Coulomb energy is larger in the case of MgReO$_3$ since the Mg$^{2+}$-O$^{2-}$ Coulomb interaction is stronger than the Li$^{+}$-O$^{2-}$ Coulomb interaction. While the phonon dispersion of $R\bar{3}c$ MgReO$_3$ displays two imaginary modes, $R3c$ MgReO$_3$ is found to be dynamically stable. The estimated interaction parameters (see Table I) of the effective Hamiltonian for MgReO$_3$ are much larger than those in LiOsO$_3$. The PTMC simulation shows that MgReO$_3$ has a FE-like phase transition at 600 K, 391 K higher than that for LiOsO$_3$. Therefore, our results suggest that MgReO$_3$ should be a room temperature "FE" metal.

Recently, Puggioni and Rondinelli[21] proposed a rule for designing a "FE" metal: The primary ingredient relies on the removal of inversion symmetry through displacements of atoms whose electronic degrees of freedom are decoupled from the states at the Fermi level. It seems that LiOsO$_3$ satisfies the above rule since Li is responsible for the FE-like distortion, while Os is responsible for the metallicity (see Fig. S1 of Supplementary Material[19]). However, the A-site driven instability in LiNbO$_3$-type system is so general that the A-site atom may be responsible for both FE-like instability and metallicity in some LiNbO$_3$-type "FE" metals. To illustrate this possibility, we consider a hypothetical system TiGaO$_3$. We find that the $R3c$ structure is more stable by 1.05 eV/f.u. than the $R\bar{3}c$ structure since there are more Ti-O bonds in $R3c$ structure. The electronic states at the Fermi level is mainly due to the 3d states of the Ti$^{3+}$ (d$^1$) ion (see Fig. S4 of Supplementary Material[19]). Therefore, in the hypothetical system TiGaO$_3$, the Ti atom is responsible for both FE-like instability and metallicity.

Finally, we will discuss the possibility of switching the two FE-like states of a "FE" metal by an electric field. If one apply an electric field to a "FE" metal through the electrodes in direct contact with the "FE" metal, the free electron in a "FE" metal will become mobile and the FE-like distortion will remain unchanged. We propose a possible way to switch the FE-like distortion of a "FE" metal by an electric field. The idea is to construct a superlattice composed by an insulating FE material and a "FE" metal. Along the superlattice direction, the whole system is not conductive due to the existence of the insulating blocks. When an electric field is applied to the superlattice, the polarization of the FE material will align along the direction of the electric field. Because of interfacial coupling (i.e., short-range pair interaction between the local modes) between the FE material and the "FE" metal, the local modes in the "FE" metal will also align along the electric field direction. We confirm this idea by examining the superlattice composed by FE insulator $LiNbO_3$ and "FE" metal $LiOsO_3$ (see Fig. 6). We find that when the "polarization" of $LiOsO_3$ is along the same direction as the polarization of $LiNbO_3$, the energy is the lowest. Therefore, the FE-like distortion of $LiOsO_3$ in this superlattice should be switchable by an electric field. In addition, we propose that it may be possible to switch the FE-like distortion of a $LiOsO_3$ slab by an electric voltage. Our DFT calculation shows that the $LiOsO_3$ slab normal to the hexagonal *c*-axis (shown in the left panel of Fig. 7) can maintain the FE-like distortion and the dipole moment along the negative *c*-axis is found to be 0.0014 e/Å per unit area. We note that the dipole moment along the *c*-axis is well-defined since the charge density is localized along the *c*-axis. This dipole moment can couple with electric field once the current flow along the *c*-axis can be suppressed. This can be achieved by leaving some interval between the electrodes and the slab (see the right panel of Fig. 7). When the distance between the electrodes and the slab is large enough so that the tunneling current is negligible, the dipole moment in the slab can be controlled by the applied electric voltage.

When we were finalizing this work, we notice that Sim and Kim[22] have already computed the phonon dispersion and electronic structure of $LiOsO_3$. Similar to their results, we find the computed structural parameters for $LiOsO_3$ are in very good agreement with the experimental values. However, the microscopic mechanism for the FE-like phase transition revealed in our work is completely new. In particular, we address why the A-site Li atom has a local polar instability and how the local modes of Li-ions interact with each other and order "ferroelectrically".

## Summary

In conclusion, the origin of the FE-like phase transition in metallic $LiOsO_3$ is revealed by performing comprehensive DFT calculations and effective Hamiltonian study. We find that the local polar instability is solely due to the fact that the A-site Li atom has a tendency to form bonds with out-of-plane O atoms. The short-range pair interactions between the local polar modes in this metallic system are found to be substantial, which help to align the local modes in a FE-like way. A route to switch the FE-like distortion of a "FE" metal by an electric field is proposed. Finally, we point out the possible existence of a new class of "FE" metal in which the same cation is responsible for both metallicity and FE-like instability, and we predict that $MgReO_3$ may be a room temperature "FE" metal, which calls for experimental confirmation.

**Acknowledgements**

Work was supported by NSFC, FANEDD, NCET-10-0351, Research Program of Shanghai Municipality and MOE, the Special Funds for Major State Basic Research, Program for Professor of Special Appointment (Eastern Scholar), and Fok Ying Tung Education Foundation.

Table I. Computed parameters of the effective Hamiltonian for LiOsO$_3$ and MgReO$_3$. The unit of $\alpha$ is eV/Å$^4$, while all the other parameters are in eV/Å$^2$.

|  | $k_2$ | $\alpha$ | $J_1$ | $J_2$ | $J_3$ | $J_4$ | $J_5$ |
|---|---|---|---|---|---|---|---|
| LiOsO$_3$ | -0.0764 | 0.6664 | -0.0194 | 0.0003 | -0.0306 | 0.0016 | 0.0063 |
| MgReO$_3$ | -0.1721 | 1.9740 | -0.0456 | 0.0141 | -0.1456 | 0.0084 | 0.0644 |

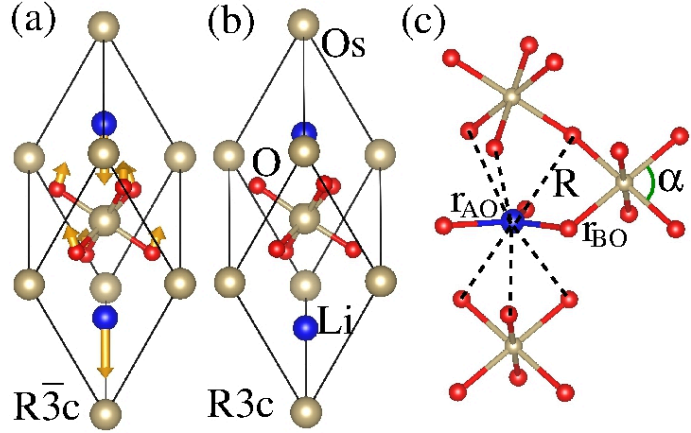

Figure 1. Atomic structures of the LiOsO$_3$ system. (a) and (b) show the $R\bar{3}c$ and $R3c$ structures, respectively. The displacements of the lowest frequency phonon mode at Γ of $R\bar{3}c$ LiOsO$_3$ are denoted by arrows (O displacements are enlarged by five times for clarity). The environment of the A-site Li atom in the $R\bar{3}c$ structure is illustrated in (c). The distance between the Li atom and the in-plane (out-of-plane) O atom is defined as $r_{AO}$ (R), while the distance between Os and O is referred as $r_{BO}$. The angle <O-Os-O is defined as α, which is very close to 90°. We prove from geometrical analysis that $R \approx \sqrt{2} r_{BO}$.

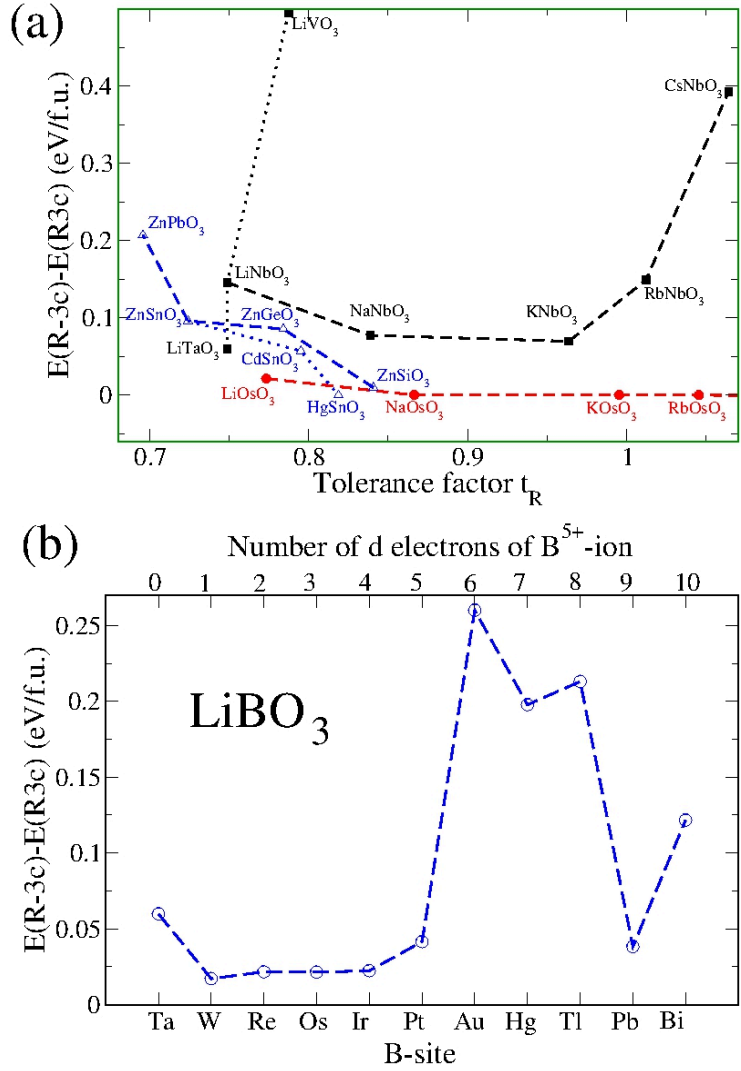

Figure 2. (a) Energy difference between the $R\bar{3}c$ and $R3c$ structures versus the tolerance factor $t_R$ (see text for the definition) for some ABO$_3$ systems. The compounds related to LiOsO$_3$, LiNbO$_3$, and ZnSnO$_3$ are connected by red, black, and blue lines, respectively. (b) Energy difference between the $R\bar{3}c$ and $R3c$ structures for some LiBO$_3$ systems where B ions are Period 6 elements of the periodic table.

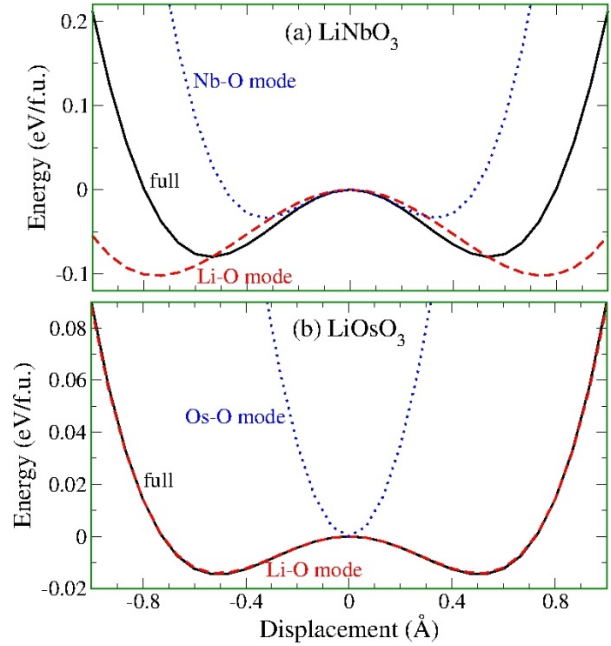

Figure 3. Potential energy versus the displacement of the lowest frequency phonon mode at Γ of the $R\bar{3}c$ structure. (a) and (b) show the case for LiNbO$_3$ and LiOsO$_3$, respectively. "Li-O mode" refers to the case where the B-site displacement of the mode is neglected, while in the case of "Nb-O mode" or "Os-O" mode, the A-site Li displacement of the mode is neglected. "full" means all displacements are taken into account.

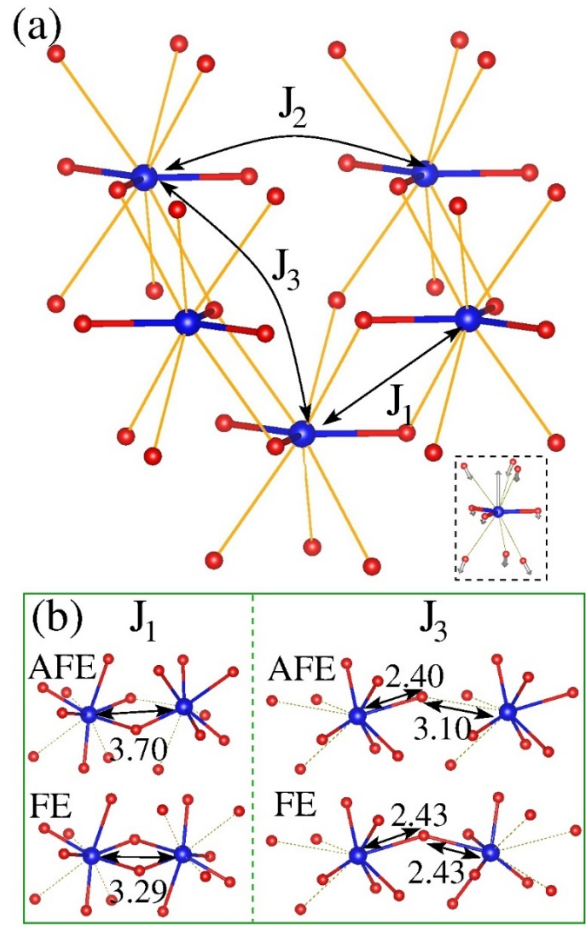

Figure 4. Local mode and the interactions between the local modes. (a) The three interaction paths between the local modes. The displacements in the local mode is displayed in the inset (O displacements are enlarged by ten times for clarity). (b) illustrates why the first ($J_1$) and third ($J_3$) interactions between the local modes are FE. It shows the atomic structures in the case of the AFE and FE aligned local dipoles (the magnitude of each local mode is 0.4 Å). The numbers denotes the distance (in Å) as indicated by the double-end arrows.

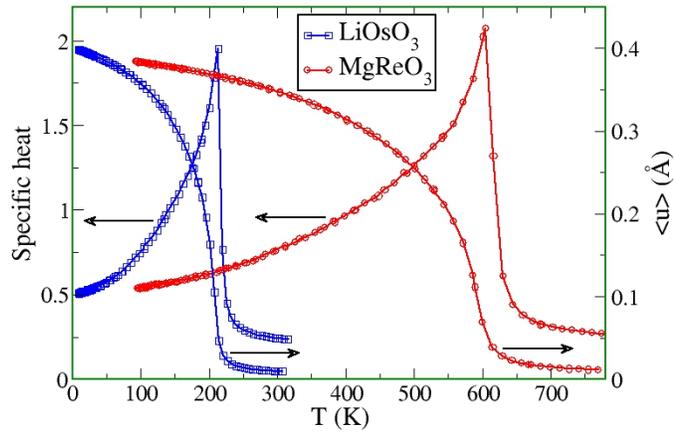

Figure 5. Finite temperature thermodynamic properties of LiOsO3 and MgReO3 from the PTMC simulations. Both the specific heat (in arbitrary unit) and order parameter (average local mode <u>) are plotted as a function of temperature.

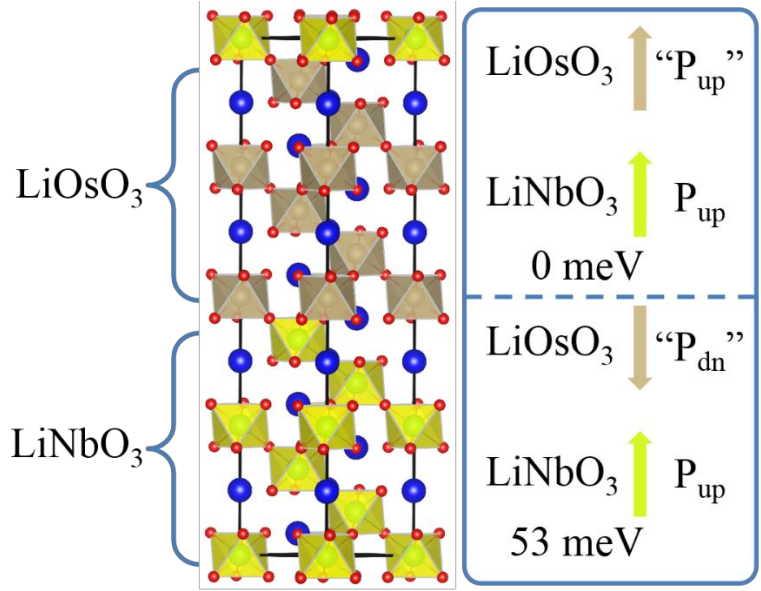

Figure 6. Structure of the superlattice composed by insulting FE LiNbO3 and "FE" metal LiOsO3. Our DFT calculation shows that the state with polarizations (P) of LiNbO3 and LiOsO3 along the same direction has a lower energy by 53 meV than that along the opposite direction.

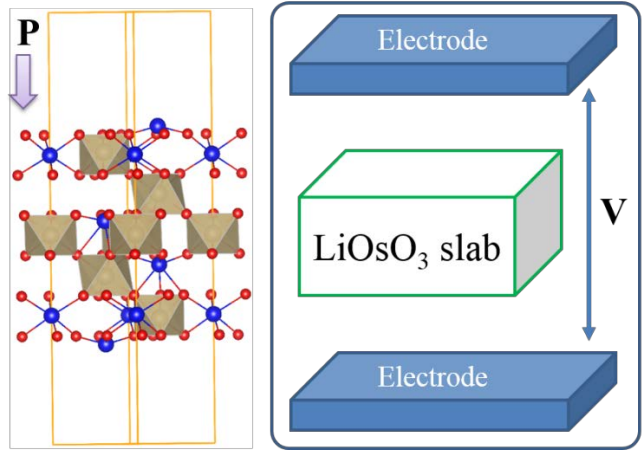

Figure 7. Schematic illustration of the switching of the FE-like distortion in a LiOsO$_3$ slab by an external electric voltage. The left panel shows the LiOsO$_3$ slab along the hexagonal $c$ axis with the dipole moment (unit area polarization P = 0.0014 e/Å) along the $-c$ direction. The right panel shows the proposed experimental set up for switching the FE-like distortion. The metallic LiOsO$_3$ slab should not be in contact with the electrodes to avoid the electric current flow between the two electrodes.